\documentclass[aps,prb,twocolumn,preprintnumbers,amsmath,amssymb,superscriptaddress]{revtex4}%

\usepackage{graphicx}
\usepackage{dcolumn}
\usepackage{amsmath}
\usepackage{color}
\usepackage{hyperref}
\usepackage{bbold}

\newcommand{\RN}[1]{\textup{\uppercase\expandafter{\romannumeral#1}}}%

\begin{document}

\title{Pressure Effect on the Spin Density Wave Transition in La$_2$PrNi$_2$O$_{6.96}$}

\author{Rustem Khasanov}
 \email{rustem.khasanov@psi.ch}
 \affiliation{PSI Center for Neutron and Muon Sciences CNM, 5232 Villigen PSI, Switzerland}

\author{Igor Plokhikh}
 \affiliation{PSI Center for Neutron and Muon Sciences CNM, 5232 Villigen PSI, Switzerland}

\author{Thomas J. Hicken}
 \affiliation{PSI Center for Neutron and Muon Sciences CNM, 5232 Villigen PSI, Switzerland}

\author{Hubertus Luetkens}
 \affiliation{PSI Center for Neutron and Muon Sciences CNM, 5232 Villigen PSI, Switzerland}

\author{Dariusz J. Gawryluk}
 \affiliation{PSI Center for Neutron and Muon Sciences CNM, 5232 Villigen PSI, Switzerland}

\author{Zurab Guguchia}
 \affiliation{PSI Center for Neutron and Muon Sciences CNM, 5232 Villigen PSI, Switzerland}

\begin{abstract}

High-pressure studies reveal a stark contrast between the superconducting properties of double-layer Ruddlesden-Popper (RP) nickelates  La$_2$PrNi$_2$O$_7$ and La$_3$Ni$_2$O$_7$. While La$_2$PrNi$_2$O$_7$ exhibits bulk superconductivity, La$_3$Ni$_2$O$_7$ displays filamentary behavior, suggesting that superconductivity is confined to phase interfaces rather than the bulk. Since magnetism emerges near the superconducting phase, understanding its differences in La$_3$Ni$_2$O$_7$ and La$_2$PrNi$_2$O$_7$ is essential for clarifying their underlying electronic and magnetic properties.
In this work we study the magnetic responce of La$_2$PrNi$_2$O$_{6.96}$ under pressures up to 2.3~GPa using the muon-spin rotation/relaxation ($\mu$SR) technique. The application of external pressure increases the N\'{e}el temperature $T_{\rm N}$ from approximately 161~K at ambient pressure ($p=0$) to about 170~K at $p=2.3$~GPa. The temperature dependence of the internal magnetic field $B_{\rm int}(T)$ ({\it i.e.} the magnetic order parameter) follows the power-law relation $B_{\rm int} = B_{\rm int}(0) \left(1 - \left[T/T_{\rm N}\right]^\alpha \right)^\beta$, with consistent exponent values of $\alpha\simeq 1.95$ and $\beta\simeq 0.35$ across different pressures.  The value of the ordered moments at the Ni sites, which is proportional to $B_{\rm int}$, remain unaffected by pressure. Our findings suggest that the magnetic properties of double-layer RP nickelate La$_3$Ni$_2$O$_7$ are broadly unaffected by Pr to La substitution.

\end{abstract}

\maketitle

\noindent \underline{\it Introduction.--} The discovery of high-temperature superconductivity in double- and triple-layered Ruddlesden-Popper (RP) nickelates, La$_{(n+1)-x}$Pr$_x$Ni$_n$O$_{3n+1}$ ($n=2$ or 3) has generated significant interest due to its implications for both fundamental physics and potential applications.\cite{Sun_Nature_2023, Zhang_NatPhys_2024, Wang_Nature_2024, Liu_NatCom_2024, Zhang_JMST_2024, Li_SciBull_2024, Li_ChinPhysLet_2024, Zhou_arxiv_2023, Zhou_arxiv_2024, Sakakibara_PRB_2024, Wang_ChinPhysLett_2024, Pei_arxiv_2024} These nickelate systems, structurally analogous to cuprates but with distinct electronic properties, offer a new platform to explore unconventional superconductivity. Notably, superconductivity under pressure was reported in these materials, with the highest superconducting transition temperature, $T_{\rm c} \simeq 80$~K, achieved in the double-layered La$_3$Ni$_2$O$_7$ system.\cite{Sun_Nature_2023, Zhang_NatPhys_2024, Wang_Nature_2024, Liu_NatCom_2024, Zhang_JMST_2024, Li_ChinPhysLet_2024, Wang_ChinPhysLett_2024, Zhou_arxiv_2023} Such $T_{\rm c}$ exceeds the boiling point of liquid nitrogen, marking an important milestone for practical superconducting applications.

Initial experiments on La$_3$Ni$_2$O$_7$ showed, however, a very small superconducting volume fraction,\cite{Zhou_arxiv_2023, Wang_PRX_2024, Li_arxiv_2025, Puphal_PRL_2024} rising a critical question of what the true superconducting phase in this material is? Addressing this question requires a comprehensive investigation into the superconducting and magnetic properties of the system.

Previous research identified two major challenges in understanding the superconductivity of La$_3$Ni$_2$O$_7$:\cite{Li_arxiv_2025} \\
(i) The formation of distinct structural forms (polymorphs), such as 2222, and 1313 within the double-layered La$_3$Ni$_2$O$_7$ system.\cite{Puphal_PRL_2024, Chen_JAChSoc_2024, Wang_InorgChem_2024, Abadi_arxiv_2024, Wang_InorgChem_2_2024} These polymorphs emerge due to variations in synthesis conditions, altering the material’s physical properties. Samples dominated by the 2222 or 1313 phases exhibited superconducting signatures around 80~K,\cite{Sun_Nature_2023, Puphal_PRL_2024} emphasizing their significance in high-temperature superconductivity research.
\\
(ii) The superconductivity in La$_3$Ni$_2$O$_7$ frequently appears in a filamentary form, with a low superconducting volume fraction.\cite{Zhou_arxiv_2023, Wang_PRX_2024, Puphal_PRL_2024} This aspect raises concerns regarding the stability of the superconducting phase. However, experimental results indicated the possibility of bulk superconductivity in this system,\cite{Li_arxiv_2024} suggesting that superconducting properties may remain stable under specific conditions, requiring further examination of the intrinsic superconducting phase.

To address the influence of structural polymorphs and the low superconducting volume fraction, recent investigations explored the partial substitution of La with Pr.\cite{Wang_Nature_2024} This substitution effectively increased the purity of the bilayer structure, favoring the formation of the 2222 phase, and resulted in an enhanced $T_c$ (up to $\simeq 83$~K) and a larger superconducting volume fraction.

A comparison of non-substituted La$_3$Ni$_2$O$_7$ with Pr-substituted La$_{3-x}$Pr$_x$Ni$_2$O$_7$ compounds is essential. In this context, the present study focuses on the magnetic properties of La$_2$PrNi$_2$O$_{6.96}$, examined using the muon-spin rotation/relaxation ($\mu\text{SR}$) technique. The results indicate that the magnetic properties of La$_2$PrNi$_2$O$_{6.96}$ under pressures up to 2.3~GPa closely match those of the non-substituted La$_3$Ni$_2$O$_7$ system. The only notable distinction is a slightly higher magnetic ordering temperature at ambient pressure: approximately 162~K for La$_2$PrNi$_2$O$_{6.96}$ compared to 152–157~K for La$_3$Ni$_2$O$_7$.\cite{Chen_PRL_2024, Khasanov_La327_arxiv_2024, Plokhikh_unp}


\noindent \underline{\it Experimental details.--} The La$_2$PrNi$_2$O$_{6.96}$ sample was obtained from the same growth batch previously used for ambient pressure studies employing neutron powder diffraction and $\mu$SR techniques.\cite{Plokhikh_unp} The sample purity and oxygen content [$7-\delta =6.96(1)$] were verified through x-ray diffraction and thermogravimetric analysis.\cite{Plokhikh_unp}

\begin{figure*}[htb]
\includegraphics[width=1\linewidth]{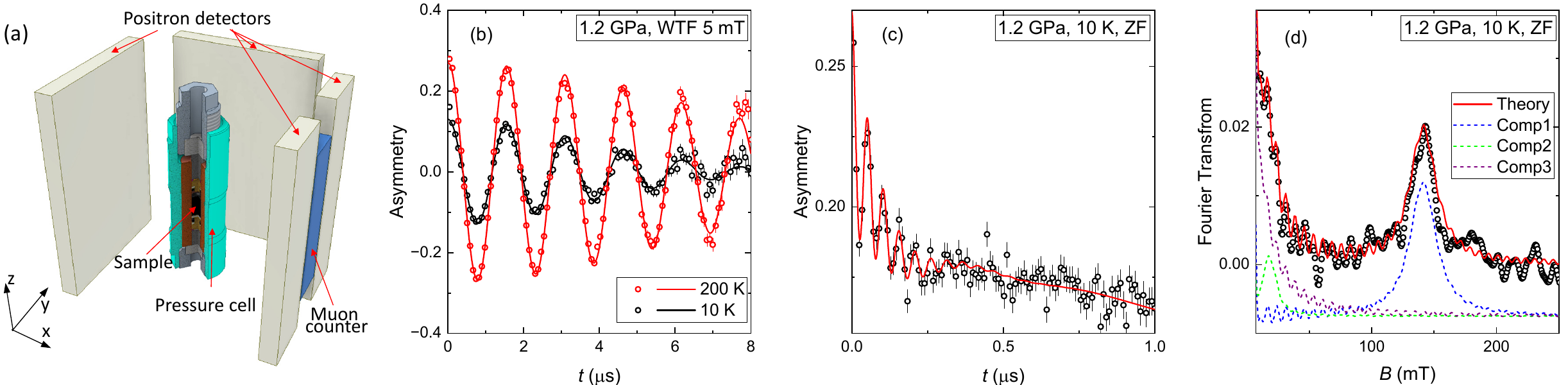}
\caption{(a) Schematic representation of the $\mu$SR experiment under pressure. The initial muon-spin direction and muon momentum are parallel to each other and antiparallel to the $x$-axis. The weak transverse field ($B_{\rm WTF} = 5$~mT), is applied parallel to the $y$-axis.
(b) WTF-$\mu$SR time-spectra of La$_2$PrNi$_2$O$_{6.96}$ recorded at $p=1.2$~GPa above ($T=200$~K) and below ($T=10$~K) the SDW transition.
(c) ZF-$\mu$SR time-spectra collected at $T=10$~K and $p=1.2$~GPa.
(d) Fourier transform of the data shown in panel (c).
The solid lines in (b)--(d) represent fits using Eq.~\ref{eq:asymmetry}, with the magnetic and nonmagnetic components described by Eqs.~\ref{eq:commensurate}, \ref{eq:GKT}, and \ref{eq:WTF}. The dashed lines in (d) represent individual internal fit components obtained from Eq.~\ref{eq:commensurate}.}
 \label{fig:experiment}
\end{figure*}

The muon-spin rotation/relaxation ($\mu$SR) experiments under pressure were conducted at the $\mu$E1 beamline using the dedicated GPD muon spectrometer (Paul Scherrer Institute, PSI Villigen, Switzerland).\cite{Khasanov_HPR_2016, Khasanov_JAP_2022} Pressures up to $\simeq 2.3$~GPa were generated using a double-wall pressure cell made of nonmagnetic MP35N alloy.\cite{Khasanov_HPR_2016} Two types of $\mu$SR experiments were performed: zero-field (ZF) and weak transverse field (WTF), with the latter applied perpendicular to the initial muon-spin polarization.

Figure~\ref{fig:experiment}~(a) presents a schematic of the $\mu$SR experiments under pressure. The sample, placed inside the pressure cell, was surrounded by positron detectors. The start of the muon clock was triggered by the muon counter. The initial muon-spin direction and muon momentum were aligned parallel to each other and antiparallel to the $x$-axis. The WTF field, $B_{\rm WTF} = 5$~mT, was applied parallel to the $y$-axis. The ZF- and WTF-$\mu$SR data were analyzed using the Musrfit package.\cite{MUSRFIT}

The data analysis was performed by separating the $\mu$SR signal into contributions from the sample (s) and the pressure cell (background, bg) responses:
\begin{equation}
A(t) = A_{\rm s}(t) + A_{\rm bg}(t) = A_{\rm s,0}P_{\rm s}(t) + A_{\rm bg,0}P_{\rm bg}(t).
\label{eq:asymmetry}
\end{equation}
Here, $A_{\rm s,0}$/$A_{\rm bg,0}$ and $P_{\rm s}(t)$/$P_{\rm bg}(t)$ represent the initial asymmetries and the muon-spin polarization functions associated with the sample and the pressure cell, respectively. The WTF and ZF contributions of the pressure cell were determined through a separate set of experiments.\cite{Khasanov_HPR_2016, Khasanov_JAP_2022}

The sample polarization function, $P_{\rm s}(t)$, was further decomposed into magnetic (m) and nonmagnetic (nm) components, with corresponding weights $f_{\rm m}$ and $1-f_{\rm m}$: $P_{\rm s}(t) = f_{\rm m}P_{\rm s,m}(t)+(1-f_{\rm m})P_{\rm s,nm}(t)$. The magnetic contribution was described by:
\begin{equation}
P_{\rm s,m}(t) = \frac{2}{3} \sum_i f_{{\rm m,}i} e^{-\lambda_{{\rm T},i} t} \cos(\gamma_\mu B_{{\rm int},i} t) + \frac{1}{3} e^{-\lambda_{\rm L}t},
\label{eq:commensurate}
\end{equation}
where $\gamma_\mu = 851.616$~MHz/T is the muon gyromagnetic ratio and $\lambda$ is the exponential relaxation rate. $B_{{\rm int},i}$, and $f_{{\rm m,}i}$ ($\sum f_{{\rm m,}i}=f_{\rm m}$) represent the internal field, and the volume fraction of the $i$-th magnetic component, respectively. The coefficients $2/3$ and $1/3$ account for powder averaging, with $2/3$ of the muon spins precessing in internal fields perpendicular (transverse, T) to the field direction and $1/3$ remaining parallel (longitudinal, L) to $B_{{\rm int},i}$. \cite{Amato-Morenzoni_book_2024, Schenck_book_1985, Yaouanc_book_2011, Blundell_book_2022}

The nonmagnetic polarization function in ZF-$\mu$SR experiments was modeled using the Gaussian Kubo-Toyabe polarization function, which is commonly applied to describe the contribution of static nuclear magnetic moments:\cite{Amato-Morenzoni_book_2024, Schenck_book_1985, Yaouanc_book_2011, Blundell_book_2022}
\begin{equation}
P_{\rm s,nm}^{\rm ZF}(t) = \frac{2}{3} (1 - \sigma_{\rm GKT}^2 t^2) \exp\left[ -\frac{\sigma_{\rm GKT}^2 t^2}{2} \right] + \frac{1}{3}.
\label{eq:GKT}
\end{equation}
Here, $\sigma_{\rm GKT}$ denotes the Gaussian Kubo-Toyabe relaxation rate.

The nonmagnetic WTF response was modeled as:
\begin{equation}
P_{\rm s,nm}^{^{\rm WTF}}(t) = e^{-\sigma_{\rm WTF}^2t^2/2}\cos(\gamma_\mu B_{\rm WTF} t +\phi),
\label{eq:WTF}
\end{equation}
where $\phi$ represents the initial phase of the muon-spin ensemble, and $\sigma_{\rm WTF}$ is the Gaussian relaxation rate of the nonmagnetic component. It should be noted that in La$_2$PrNi$_2$O$_{7-\delta}$, the magnetic contribution to the muon-spin polarization function [$P_{\rm s,m}(t)$, Eq.~\ref{eq:commensurate}], nearly vanishes for $t\gtrsim 0.3$~$\mu$s.\cite{Plokhikh_unp, Chen_arxiv_2024} Consequently, the analysis of WTF data was performed for $t > 0.3$~$\mu$s, where only the nonmagnetic term described by Eq.~\ref{eq:WTF} remains.


\noindent \underline{\it Experimental data.--} Figure~\ref{fig:experiment}~(b) presents the WTF-$\mu$SR time-spectra collected above ($T=200$~K) and below ($T=10$~K) the magnetic transition temperature ($T_{\rm N} \sim 160$~K) at a pressure of $p=1.2$~GPa. The reduction in asymmetry at $T=10$~K results from the emergence of spin-density wave (SDW) order. The solid lines represent fits using Eq.~\ref{eq:asymmetry}, with the nonmagnetic polarization function described by Eq.~\ref{eq:WTF}. The remaining oscillations at $T=10$~K correspond to the background contribution from muons stopping in the pressure cell walls ($\simeq 50$\% in our case).

The ZF response of La$_2$PrNi$_2$O$_{6.96}$, measured at $T=10$~K and $p=1.2$~GPa, is shown in Fig.~\ref{fig:experiment}~(c). The observed oscillations originate from the precession of muon spin polarization in internal fields caused by the ordered magnetic moments at the muon stopping sites.\cite{Amato-Morenzoni_book_2024, Schenck_book_1985, Yaouanc_book_2011, Blundell_book_2022} The Fourier transform in Fig.~\ref{fig:experiment}~(d) reveals an internal field distribution consisting of two oscillating components and one fast-relaxing component.
It should be noted that the presence of a $\simeq 50$\% background contribution from muons stopping in the pressure cell walls, combined with the lower counting rate of the GPD spectrometer compared to the low-background Flame muon instrument (PSI Villigen, Switzerland), prevented the resolution of two low-intensity peaks at $B_{\rm int} \simeq 100$~mT and 50~mT.\cite{Plokhikh_unp} These peaks were associated in Ref.~\onlinecite{Plokhikh_unp} with the presence of a $\sim 10$\% 1313 structural phase in the La$_2$PrNi$_2$O$_{6.96}$ sample. Consequently, the fit of the ZF-$\mu$SR data was performed by incorporating three magnetic components in Eq.~\ref{eq:commensurate}: a fast-oscillating, a slow-oscillating, and a fast-decaying component, as represented by the dashed curves in Fig.~\ref{fig:experiment}~(c).

\begin{figure}[htb]
\includegraphics[width=0.7\linewidth]{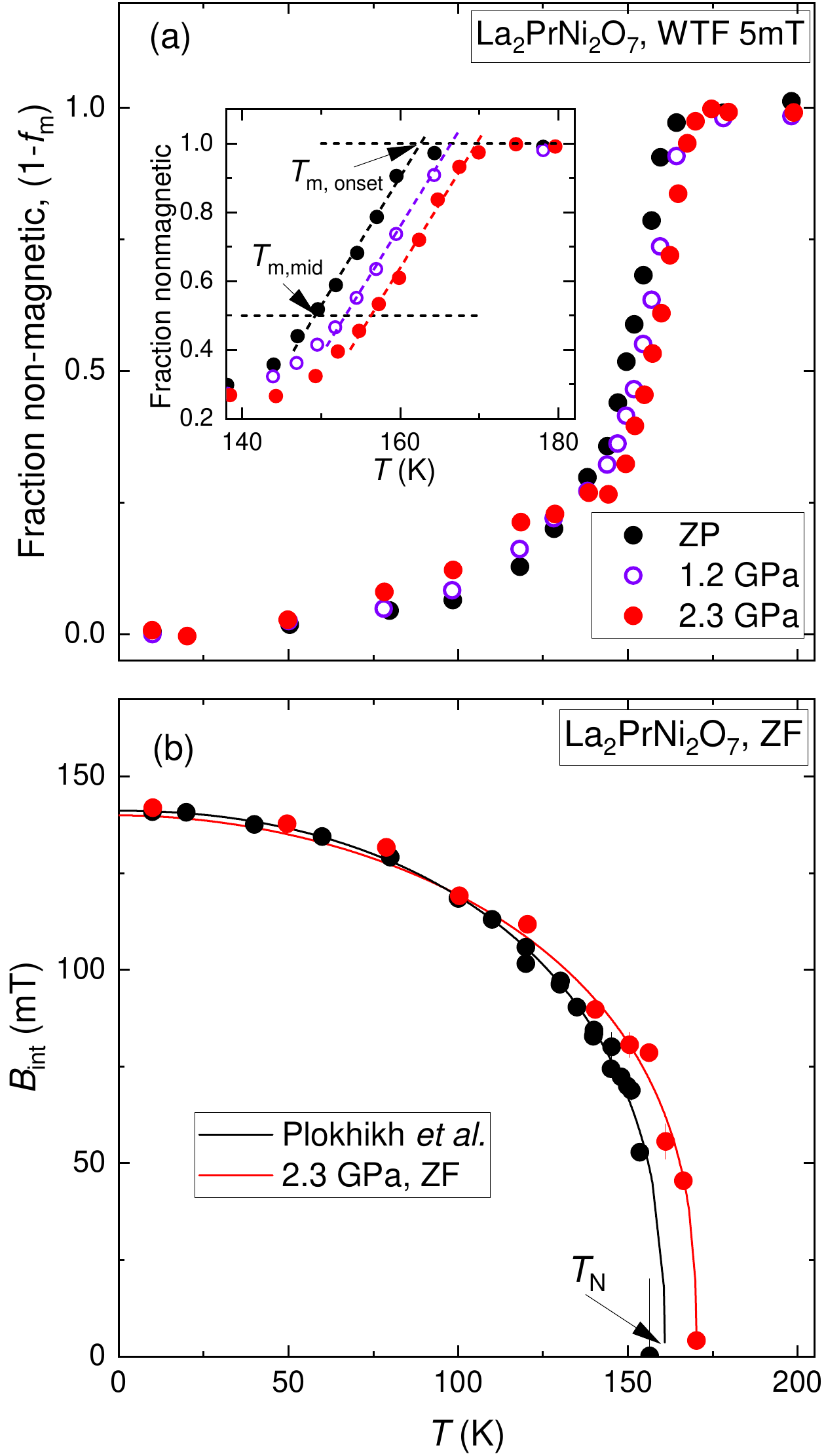}
\caption{
(a) Temperature dependence of the nonmagnetic volume fraction ($1-f_{\rm m}$) obtained from WTF-$\mu$SR experiments at pressures $p=0.0$, 1.2, and 2.3~GPa. The inset presents an extended view of the ($1-f_{\rm m}$) vs. $T$ data. The magnetic transition temperatures, $T_{\rm N,mid}$ and $T_{\rm N,onset}$, are determined as the crossing points of the linear fits of ($1-f_{\rm m}$) vs. $T$ in the vicinity of the magnetic transition with the reference lines $1-f_{\rm m}=0.5$ and 1.0, respectively.
(b) Temperature dependence of the fast-oscillating component of the internal field $B_{\rm int}$, measured at $p=0.0$ and 2.3~GPa. The ambient pressure data are taken from Ref.~\onlinecite{Plokhikh_unp}. The solid lines represent power-law fits using Eq.~\ref{eq:power-law}, with the corresponding parameters summarized in Table~\ref{table:Results}.}
 \label{fig:WTF-ZF_results}
\end{figure}

Figure~\ref{fig:WTF-ZF_results} presents the temperature dependence of various parameters obtained from fits of WTF- and ZF-$\mu$SR data. Following Fig.~\ref{fig:WTF-ZF_results}~(a), the La$_2$PrNi$_2$O$_{6.96}$ sample exhibits purely magnetic behavior at low temperatures and transitions to a nonmagnetic state at temperatures exceeding $\sim 150$~K. The extended region of the $1-f_{\rm m}$ vs. $T$ curves, displayed in the inset, reveals that the magnetic transition temperatures, determined as $T_{\rm m, mid}$ and $T_{\rm m, onset}$, increase with applied pressure.

The temperature dependence of the fast precessing component $B_{\rm int,fast}$, measured at $p=2.3$~GPa, is shown in Fig.~\ref{fig:WTF-ZF_results}~(b). Hereafter, only the strongest ({\it i.e.}, fast precessing) component is considered, and the subscript `fast' in $B_{\rm int,fast}$ is omitted. For comparison, Fig.~\ref{fig:WTF-ZF_results} also includes the $B_{\rm int}(T)$ dependence obtained at ambient pressure for a sample from the same growth batch, as reported in Ref.~\onlinecite{Plokhikh_unp}. The solid lines represent fits using the power-law function:\cite{Pratt_JPCM_2007, Sugiyama_PRB_2009, Maisuradze_PRB_2011}
\begin{equation}
B_{\rm int}(T)= B_{\rm int}(0)[1-(T/T_{\rm N})^\alpha]^\beta,
 \label{eq:power-law}
\end{equation}
where $T_{\rm N}$ denotes the N\'{e}el temperature, and $B_{\rm int}(0)$ is the zero-temperature value of the internal field. Here we distinguish between the N\'{e}el temperature $T_{\rm N}$, which corresponds to the vanishing of the magnetic order parameter [$B_{\rm int}\rightarrow 0$, Fig.~\ref{fig:WTF-ZF_results}~(b)], and the transition temperatures $T_{\rm m, onset}$ and $T_{\rm m, mid}$, as determined from changes in the magnetic volume fraction $f_{\rm m}$ [the inset of Fig.~\ref{fig:WTF-ZF_results}~(a)].

\begin{figure}[htb]
\includegraphics[width=0.7\linewidth]{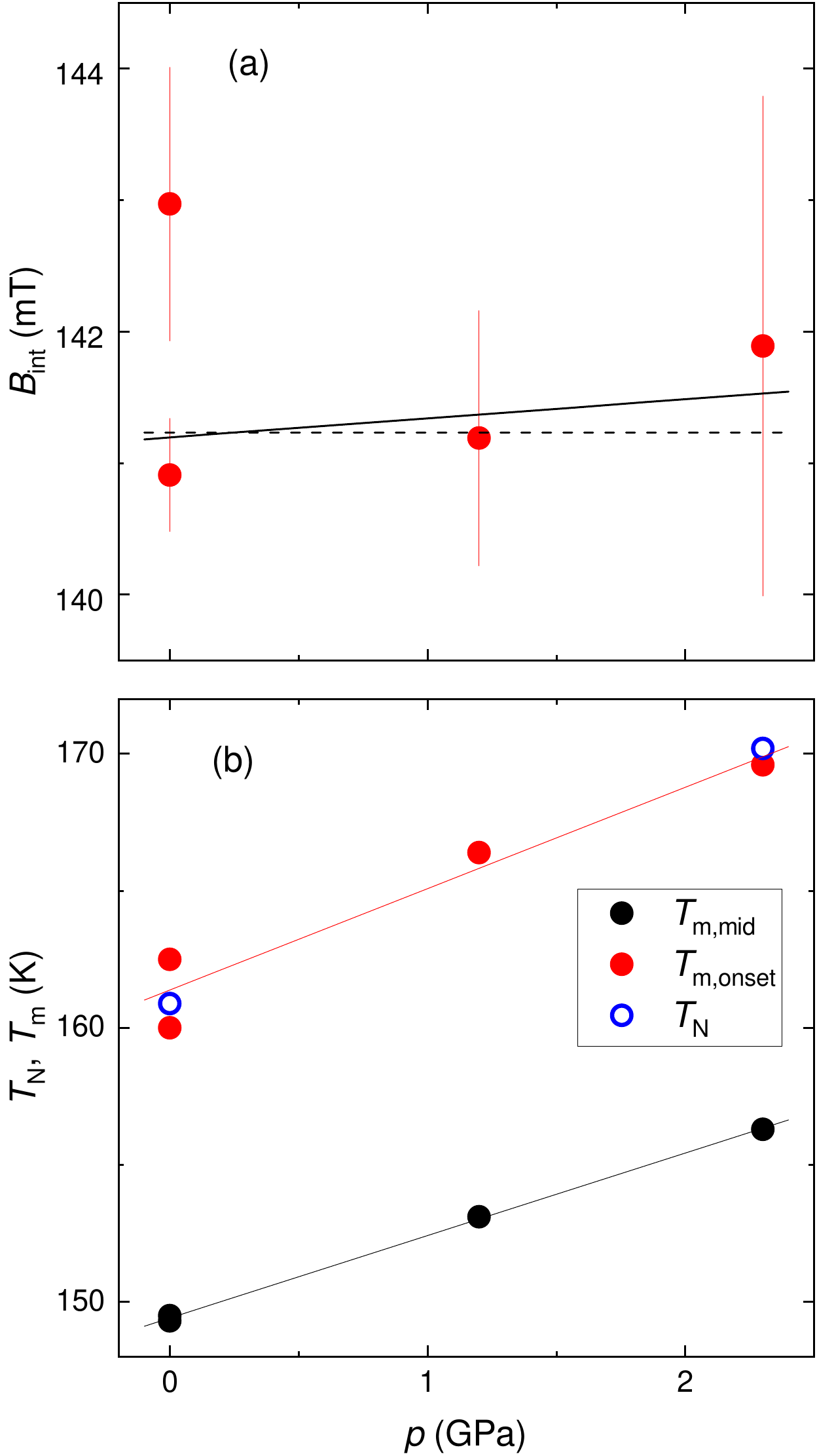}
\caption{
(a) Pressure dependence of the internal magnetic field $B_{\rm int}$ at $T=10$~K.
(b) Pressure dependencies of magnetic transition temperatures as determined from analysis of WTF- and ZF-$\mu$SR data.
The solid lines in (a) and (b) are linear fits.}
 \label{fig:Tn-Bint}
\end{figure}

The magnetic transition temperatures ($T_{\rm m, mid}$, $T_{\rm m, onset}$, and $T_{\rm N}$), the internal fields at $T=0$ and $T=10$~K [$B_{\rm int}(0)$ and $B_{\rm int}(10{\rm ~K})$], and the exponents ($\alpha$ and $\beta$) are summarized in Table~\ref{table:Results}. The pressure dependencies of the transition temperatures and the internal field are presented in Fig.~\ref{fig:Tn-Bint}. Additionally, the ambient pressure data points from Ref.~\onlinecite{Plokhikh_unp} are included for comparison.

\begin{table*}[htb]
\caption{The magnetic transition temperatures ($T_{\rm m, mid}$, $T_{\rm m, onset}$, and $T_{\rm N}$), the internal fields at $T=0$ and 10~K [$B_{\rm int}(0)$ and $B_{\rm int}(10{\rm ~K})$]  and the exponents ($\alpha$ and $\beta$) as obtained from fits of WTF- and ZF-$\mu$SR data of La$_2$PrNi$_2$O$_{6.96}$. }
\begin{tabular}{cccccccc}
\hline
\hline
Presure & $T_{\rm m, mid}$ & $T_{\rm m, onset}$ & $T_{\rm N}$&$B_{\rm int}(10{\rm ~K})$&$B_{\rm int}(0)$ &$\alpha$&$\beta$ \\
(GPa  ) & (K)              & (K)                & (K)        & (mT)             &        &         \\
\hline 
0.0 & 149.3(3) & 162.5(3) & --      & 143.0(1.0)         &--       &-- &--\\
1.2 & 153.1(3) & 166.4(4) & --      & 141.2(1.0)         & --      & --&--\\
2.3 & 156.3(3) & 169.6(4) &170.2(4) & 141.9(1.9)& 139.9(3.5)&1.91(27)& 0.35(2)\\
\\
0.0\footnotemark[1] &149.5(2)\footnotemark[1]& 162.0(2)\footnotemark[1]& 160.9(5)\footnotemark[1]&140.9(4)\footnotemark[1]& 141.4(6)\footnotemark[1]&2.02(15)\footnotemark[1]&0.35(2)\footnotemark[1] \\
\hline
\hline
\end{tabular}
\footnotetext[1]{After Ref.~\onlinecite{Plokhikh_unp}}
\label{table:Results} 
\end{table*}

\noindent \underline{\it Discussions.--} From the results presented in Figs.~\ref{fig:experiment}, \ref{fig:WTF-ZF_results}, \ref{fig:Tn-Bint}, and Table~\ref{table:Results}, four key observations emerge: \\
(i) The magnetic field distribution is well described by the sum of Lorentzian peaks [Figs.~\ref{fig:experiment}~(c) and (d)], suggesting a commensurate type of magnetic order in the La$_{2}$PrNi$_2$O$_{6.96}$ sample studied here. This result is consistent with ZF-$\mu$SR data obtained by Chen {\it et al.} for La$_{1.9}$Pr$_{1.1}$Ni$_2$O$_{6.97}$ and La$_{3}$Ni$_2$O$_{6.92}$ (Refs.~\onlinecite{Chen_arxiv_2024} and \onlinecite{Chen_PRL_2024}), as well as our own studies of La$_{3}$Ni$_2$O$_{7-\delta}$ and La$_{3}$Ni$_2$O$_{6.99}$ (Refs.~\onlinecite{Khasanov_La327_arxiv_2024} and \onlinecite{Plokhikh_unp}). At the same time, this differs from the incommensurate magnetic structure reported for the $n=3$ Ruddlesden-Popper nickelate La$_4$Ni$_3$O$_{10}$.\cite{Zhang_NatCom_2020, Khasanov_La4310_unp} \\
(ii) The temperature dependence of the internal magnetic field, measured at $p=0.0$ and 2.3~GPa, is well described by the power-law function [Eq.~\ref{eq:power-law} and Fig.~\ref{fig:WTF-ZF_results}~(b)]. The exponent values $\alpha\simeq1.95$ and $\beta\simeq0.35$ remain nearly unchanged between ambient- and high-pressure measurements (Table~\ref{table:Results}). This suggests that pressure, at least up to $p\simeq 2.3$~GPa (the highest pressure reached in this study), does not affect the type of magnetic order, indicating that La$_{2}$PrNi$_2$O$_{6.96}$ remains close to a 3D magnet, which is characterized by $\beta\simeq1/3$.
At present, it is not possible to distinguish between the 3D Heisenberg model ($\beta=0.365$), the 3D Ising model ($\beta=0.325$), or the 3D XY model ($\beta=0.345$).\cite{Kaul_JMMM_1985, Guillou_PRB_1985}\\
(iii) The internal field $B_{\rm int}$(10~K) remains nearly independent of pressure [Fig.~\ref{fig:Tn-Bint}~(a)]. Linear fits with a free slope and a slope fixed to zero yield $B_{\rm int}(p)=141.2(5){\rm ~mT} + p\cdot0.14(82){\rm ~mT/GPa}$ and $B_{\rm int}(p)=141.2(4)$~mT, respectively. The pressure independence of $B_{\rm int}$ and its low-temperature values are consistent with those reported for the non-substituted La$_{3}$Ni$_2$O$_{7-\delta}$ system [$B_{\rm int}(0,p)\simeq 143$~mT, Ref.~\onlinecite{Khasanov_La327_arxiv_2024}].  Given the similarity between the magnetic structures of La$_3$Ni$_2$O$_7$ and La$_2$PrNi$_2$O$_7$, as determined by neutron powder diffraction and ZF-$\mu$SR experiments,\cite{Plokhikh_unp} and considering the proportionality between the internal field and the ordered Ni moments ($B_{\rm int}\propto m_{\rm Ni}$), these results suggest that not only the magnetic structures but also the values of the ordered magnetic moments remain unchanged in La$_{3-x}$Pr$_x$Ni$_2$O$_{7}$ (at least for $x\lesssim 1.0$ and $p\lesssim 2.3$~GPa).\\
(iv) The magnetic ordering temperatures, as determined from WTF- and ZF-$\mu$SR experiments, increase with pressure [Fig.~\ref{fig:Tn-Bint}~(b)]. Interestingly, the N\'{e}el temperature $T_{\rm N}$, defined as the temperature where $B_{\rm int}$ vanishes on increasing $T$ in ZF-$\mu$SR studies, coincides with the onset of magnetism observed in WTF experiments, {\it i.e.}, $T_{\rm N}\simeq T_{\rm m, onset}>T_{\rm m, mid}$. Typically, $T_{\rm N}$ is lower than the magnetism onset temperature due to magnetic fluctuations, which remain detectable above the N\'{e}el temperature within the muon time window (typically $10^4-10^{12}$~Hz, Ref.~\onlinecite{Amato-Morenzoni_book_2024, Blundell_book_2022}). At present, no clear explanation for this observation is available.  Linear fits to the data in Fig.~\ref{fig:Tn-Bint}~(b) yield
$T_{\rm N}\simeq T_{\rm m,onset}=161.3(9){\rm ~K}+p\cdot3.7(7){\rm ~K/GPa}$ and $T_{\rm m,mid}=149.4(7){\rm K}+p\cdot 3.0(6){\rm ~K/GPa}$. These pressure coefficients are comparable with ${\rm d}T_{\rm N}/{\rm d}p=2.8(3)$~K/GPa, as reported for non-substituted La$_3$Ni$_2$O$_{7-\delta}$ in Ref.~\onlinecite{Khasanov_La327_arxiv_2024}.

Overall, our results indicate that the partial substitution of La with Pr in the double-layer RP nickelate system La$_{3-x}$Pr$_x$Ni$_2$O$_7$ does not introduce significant changes in the spin-density wave (SDW) order. The only observed difference is a slightly higher magnetic ordering temperature at ambient pressure: 162~K for La$_2$PrNi$_2$O$_7$ compared to 152–157~K for La$_3$Ni$_2$O$_7$.\cite{Chen_PRL_2024, Khasanov_La327_arxiv_2024, Plokhikh_unp}
If superconducting pairing is driven by magnetic interactions,\cite{Liu_PRL_2023, Qin_PRB_2023, Sakakibara_PRL_2024_2, Shen_ChinPhysLett_2023, Xue_ChiPhysLett_2024, Yang_PRB_2023, Yang_PRB_2023_2, Zhang_PRB_2023, Luo_npjQuant_2024, Fan_PRB_2024, Heier_PRB_2024, Jiang_ChinPhysLett_2024, Lechermann_PRB_2023, Boetzel_PRB_2024} no significant changes in the superconducting response should be expected as a result of La substitution by Pr in La$_{3-x}$Pr$_x$Ni$_2$O$_7$ compounds.


\noindent \underline{\it Conclusions.--} This work presents a detailed investigation of the magnetic properties of La$_2$PrNi$_2$O$_{6.96}$ under varying pressures, providing valuable insights into the interplay between chemical substitution, compression, and magnetic ordering in this system. A key finding is that the replacement of La with Pr has only a minor influence on the magnetic properties, with the ordered magnetic moments remaining constant across the studied pressure range.

The observed increase in the magnetic ordering temperature $T_{\rm N}$, from $\simeq 161$~K at ambient pressure to $\simeq 170$~K at $p=2.3$~GPa, underscores the moderate but measurable impact of external pressure on the system’s magnetic behavior. The consistency in the power-law parameters describing the temperature dependence of $B_{\rm int}$ further highlights the robustness of magnetic ordering under pressure. Notably, the results are in good agreement with observations for the unsubstituted La$_3$Ni$_2$O$_7$ system, reinforcing the conclusion that the partial substitution of La with Pr does not introduce significant magnetic anomalies. The slightly elevated N\'{e}el temperature $T_{\rm N}$ at ambient pressure for La$_2$PrNi$_2$O$_7$ compared to La$_3$Ni$_2$O$_7$ remains the most notable distinction, suggesting subtle effects arising from Pr substitution.



\begin{thebibliography}{99}

\bibitem{Sun_Nature_2023} Hualei Sun, Mengwu Huo, Xunwu Hu, Jingyuan Li, Zengjia Liu, Yifeng Han, Lingyun Tang, Zhongquan Mao, Pengtao Yang, Bosen Wang, Jinguang Cheng, Dao-Xin Yao, Guang-Ming Zhang, and Meng Wang,  {\it Signatures of superconductivity near 80~K in a nickelate under high pressure}, Nature {\bf 621}, 493 (2023).\\
    \url{https://doi.org/10.1038/s41586-023-06408-7}

\bibitem{Zhang_NatPhys_2024} Yanan Zhang, Dajun Su, Yanen Huang, Zhaoyang Shan, Hualei Sun, Mengwu Huo, Kaixin Ye, Jiawen Zhang, Zihan Yang, Yongkang Xu, Yi Su, Rui Li, Michael Smidman, Meng Wang, Lin Jiao, and  Huiqiu Yuan, {\it High-temperature superconductivity with zero resistance and strange-metal behaviour in La$_3$Ni$_2$O$_{7-\delta}$}, Nat. Phys. {\bf 20}, 1269, (2024).\\
    \url{https://doi.org/10.1038/s41567-024-02515-y}

\bibitem{Wang_Nature_2024} Ningning Wang, Gang Wang, Xiaoling Shen, Jun Hou, Jun Luo, Xiaoping Ma, Huaixin Yang, Lifen Shi, Jie Dou, Jie Feng, Jie Yang, Yunqing Shi, Zhian Ren, Hanming Ma, Pengtao Yang, Ziyi Liu, Yue Liu, Hua Zhang, Xiaoli Dong, Yuxin Wang, Kun Jiang, Jiangping Hu, Shoko Nagasaki, Kentaro Kitagawa, Stuart Calder, Jiaqiang Yan, Jianping Sun, Bosen Wang, Rui Zhou, Yoshiya Uwatoko, and Jinguang Cheng, {\it Bulk high-temperature superconductivity in pressurized tetragonal La$_2$PrNi$_2$O$_7$}, Nature {\bf 634}, 579 (2024).\\
    \url{https://doi.org/10.1038/s41586-024-07996-8}

\bibitem{Liu_NatCom_2024} Zhe Liu, Mengwu Huo, Jie Li, Qing Li, Yuecong Liu, Yaomin Dai, Xiaoxiang Zhou, Jiahao Hao, Yi Lu, Meng Wang, and Hai-Hu Wen, {\it Electronic correlations and energy gap in the bilayer nickelate La$_3$Ni$_2$O$_7$}, Nat. Commun. {\bf 15}, 7570 (2024).\\
    \url{https://doi.org/10.1038/s41467-024-52001-5}

\bibitem{Zhang_JMST_2024} Mingxin Zhang, Cuiying Pei, Qi Wang, Yi Zhao, Changhua Li, Weizheng Cao, Shihao Zhu, Juefei Wu, and Yanpeng Qi, {\it Effects of pressure and doping on Ruddlesden-Popper phases La$_{n+1}$Ni$_n$O$_{3n+1}$}, J. Mater. Sci. Technol. {\bf 185}, 147, (2024).\\
    \url{https://doi.org/10.1016/j.jmst.2023.11.011}

\bibitem{Li_ChinPhysLet_2024} Yidian Li, Xian Du, Yantao Cao, Cuiying Pei, Mingxin Zhang, Wenxuan Zhao, Kaiyi Zhai, Runzhe Xu, Zhongkai Liu, Zhiwei Li, Jinkui Zhao, Gang Li, Yanpeng Qi, Hanjie Guo, Yulin Chen, and Lexian Yang, {\it Electronic Correlation and Pseudogap-Like Behavior of High-Temperature Superconductor La$_3$Ni$_2$O$_7$}, Chinese Phys. Lett. {\bf 41}, 087402, (2024).\\
    \url{https://doi.org/10.1088/0256-307X/41/8/087402}

\bibitem{Wang_ChinPhysLett_2024} Meng Wang, Hai-Hu Wen, Tao Wu, Dao-Xin Yao, and Tao Xiang, {\it  Normal and superconducting properties of La$_3$Ni$_2$O$_7$},  Chinese Phys. Lett. {\bf 41}, 077402, (2024).\\
    \url{https://doi.org/10.1088/0256-307X/41/7/077402}

\bibitem{Zhou_arxiv_2023} Yazhou Zhou, Jing Guo, Shu Cai, Hualei Sun, Pengyu Wang, Jinyu Zhao, Jinyu Han, Xintian Chen, Yongjin Chen, Qi Wu, Yang Ding, Tao Xiang, Ho-kwang Mao, Liling Sun {\it Investigations of key issues on the reproducibility of high Tc superconductivity emerging from compressed La$_3$Ni$_2$O$_7$}, Matter Radiat. Extremes {\bf 10}, 027801 (2025)\\
    \url{https://doi.org/10.1063/5.0247684}




\bibitem{Li_SciBull_2024} Yidian Li, Yantao Cao, Liangyang Liu, Pai Peng, Hao Lin, Cuiying Pei, Mingxin Zhang, Heng Wu, Xian Du, Wenxuan Zhao, Kaiyi Zhai, Zhang Xuefeng, Jinkui Zhao, Miaoling Lin, Pingheng Tan, Yanpeng Qi, Gang Li, Guo Hanjie, Luyi Yang, Lexian Yang, {\it Distinct ultrafast dynamics of bilayer and trilayer nickelate superconductors regarding the density-wave-like transitions}, Sci. Bull. {\bf 70}, 180 (2024).\\
    \url{https://doi.org/10.1016/j.scib.2024.10.011}

\bibitem{Zhou_arxiv_2024} Xiaoxiang Zhou, Weihong He, Zijian Zhou, Kaipeng Ni, Mengwu Huo, Deyuan Hu, Yinghao Zhu, Enkang Zhang, Zhicheng Jiang, Shuaikang Zhang, Shiwu Su, Juan Jiang, Yajun Yan, Yilin Wang, Dawei Shen, Xue Liu, Jun Zhao, Meng Wang, Mengkun Liu, Zengyi Du, and Donglai Feng, {\it Revealing nanoscale structural phase separation in La$_3$Ni$_2$O$_{7-delta}$  single crystal via scanning near field optical microscopy}, arXiv:2410.06602 (2024).\\
    \url{https://doi.org/10.48550/arXiv.2410.06602}


\bibitem{Sakakibara_PRB_2024} Hirofumi Sakakibara, Masayuki Ochi, Hibiki Nagata, Yuta Ueki, Hiroya Sakurai, Ryo Matsumoto, Kensei Terashima, Keisuke Hirose, Hiroto Ohta, Masaki Kato, Yoshihiko Takano, and Kazuhiko Kuroki, {\it Theoretical analysis on the possibility of superconductivity in the trilayer Ruddlesden-Popper nickelate La$_4$Ni$_3$O$_{10}$ under pressure and its experimental examination: Comparison with La$_3$Ni$_2$O$_7$}, Phys. Rev. B {\bf 109}, 144511, (2024).\\
    \url{https://doi.org/10.1103/PhysRevB.109.144511}

\bibitem{Pei_arxiv_2024} Cuiying Pei, Mingxin Zhang, Di Peng, Shangxiong Huangfu, Shihao Zhu, Qi Wang, Juefei Wu, Zhenfang Xing, Lili Zhang, Yulin Chen, Jinkui Zhao, Wenge Yang, Hongli Suo, Hanjie Guo, Qiaoshi Zeng, Yanpeng Qi, {\it Pressure-Induced Superconductivity in Pr$_4$Ni$_3$O$_{10}$ Single Crystals}, arXiv:2411.08677v1. \\
    \url{https://doi.org/10.48550/arXiv.2411.08677}


\bibitem{Wang_PRX_2024} G. Wang, N. N. Wang, X. L. Shen, J. Hou, L. Ma, L. F. Shi, Z. A. Ren, Y. D. Gu, H. M. Ma, P. T. Yang, Z. Y. Liu, H. Z. Guo, J. P. Sun, G. M. Zhang, S. Calder, J.-Q. Yan, B. S. Wang, Y. Uwatoko, and J.-G. Cheng, {\it Pressure-Induced Superconductivity In Polycrystalline La$_3$Ni$_2$O$_{7-\delta}$}, Phys. Rev. X {\bf 14}, 011040 (2024).\\
    \url{https://doi.org/10.1103/PhysRevX.14.011040}

\bibitem{Li_arxiv_2025} Feiyu Li, Yinqiao Hao, Ning Guo, Jian Zhang, Qiang Zheng, Guangtao Liu, and Junjie Zhang, {\it Signature of superconductivity in pressurized La$_4$Ni$_3$O$_{10-x}$ single crystals grown at ambient pressure}, arXiv:2501.13511v1 (2025). \\
    \url{https://doi.org/10.48550/arXiv.2501.13511}

\bibitem{Puphal_PRL_2024} P. Puphal, P. Reiss, N. Enderlein, Y.-M. Wu, G. Khaliullin, V. Sundaramurthy, T. Priessnitz, M. Knauft, A. Suthar, L. Richter, M. Isobe, P. A. van Aken, H. Takagi, B. Keimer, Y. E. Suyolcu, B. Wehinger, P. Hansmann, and M. Hepting, {\it Unconventional Crystal Structure of the High-Pressure Superconductor La$_3$Ni$_2$O$_7$}, Phys. Rev. Lett. {\bf 133}, 146002 (2024). \\
    \url{https://doi.org/10.1103/PhysRevLett.133.146002}

\bibitem{Chen_JAChSoc_2024} Xinglong Chen, Junjie Zhang, Amandeep S. Thind, Shashank Sharma, Harrison LaBollita, Gregory Peterson, Hangwen Zheng, Daniel Phelan, Antia S. Botana, Robert F. Klie, and John F. Mitchell,  {\it Polymorphism in the Ruddlesden–Popper Nickelate La$_3$Ni$_2$O$_7$: Discovery of a Hidden Phase with Distinctive Layer Stacking}, J. Am. Chem. Soc. {\bf 146}, 3640 (2024).\\
    \url{https://doi.org/10.1021/jacs.3c14052}

\bibitem{Wang_InorgChem_2024} Haozhe Wang, Long Chen, Aya Rutherford, Haidong Zhou, and Weiwei Xie, {\it Long-Range Structural Order in a Hidden Phase of Ruddlesden–Popper Bilayer Nickelate La$_3$Ni$_2$O$_7$}, Inorg. Chem. {\bf 63}, 5020 (2024).\\
     \url{https://doi.org/10.1021/acs.inorgchem.3c04474}


\bibitem{Abadi_arxiv_2024} Sebastien N. Abadi, Ke-Jun Xu, Eder G. Lomeli, Pascal Puphal, Masahiko Isobe, Yong Zhong, Alexei V. Fedorov, Sung-Kwan Mo, Makoto Hashimoto, Dong-Hui Lu, Brian Moritz, Bernhard Keimer, Thomas P. Devereaux, Matthias Hepting, and Zhi-Xun Shen, {\it Electronic Structure of the Alternating Monolayer-Trilayer Phase of La$_3$Ni$_2$O$_7$}, arXiv:2402.07143v2 (2024).\\
    \url{https://doi.org/10.48550/arXiv.2402.07143}

\bibitem{Wang_InorgChem_2_2024} Haozhe Wang, Haidong Zhou, Weiwei Xie, {\it Temperature-Dependent Structural Evolution of Ruddlesden–Popper Bilayer Nickelate La$_3$Ni$_2$O$_7$}, Inorg. Chem. {\bf 64}, 828 (2024).\\
    \url{https://doi.org/10.1021/acs.inorgchem.4c03042}


\bibitem{Li_arxiv_2024} Jingyuan Li, Peiyue Ma, Hengyuan Zhang, Xing Huang, Chaoxin Huang, Mengwu Huo, Deyuan Hu, Zixian Dong, Chengliang He, Jiahui Liao, Xiang Chen, Tao Xie, Hualei Sun, and Meng Wang, {\it Pressure-driven right-triangle shape superconductivity in bilayer nickelate La$_3$Ni$_2$O$_7$},  arXiv:2404.11369v2 (2024).\\
    \url{https://doi.org/10.48550/arXiv.2404.11369}

\bibitem{Plokhikh_unp} Igor Plokhikh {\it et al.}, submitted for publication.

\bibitem{Chen_PRL_2024} Kaiwen Chen, Xiangqi Liu, Jiachen Jiao, Muyuan Zou, Chengyu Jiang, Xin Li, Yixuan Luo, Qiong Wu, Ningyuan
    Zhang, Yanfeng Guo, and Lei Shu, {\it Evidence of Spin Density Waves in La$_3$Ni$_2$O$_{7-\delta}$}, Phys. Rev. Lett. {\bf 132}, 256503 (2024). \\
    \url{https://doi.org/10.1103/PhysRevLett.132.256503}

\bibitem{Khasanov_La327_arxiv_2024} Rustem Khasanov, Thomas J. Hicken, Dariusz J. Gawryluk, Vahid Sazgari, Igor Plokhikh, Lo\"{\i}c Pierre Sorel, Marek Bartkowiak, Steffen B\"{o}tzel, Frank Lechermann, Ilya M. Eremin, Hubertus Luetkens, and Zurab Guguchia,  {\it Pressure-enhanced splitting of density wave transitions in  La3Ni2O$_{7-\delta}$},  Nat. Phys. (2025). \\ \url{https://doi.org/10.1038/s41567-024-02754-z}

\bibitem{Khasanov_HPR_2016} R. Khasanov, Z. Guguchia, A. Maisuradze, D. Andreica, M. Elender, A. Raselli, Z. Shermadini, T. Goko, F. Knecht, E. Morenzoni, and A. Amato, {\it High pressure research using muons at the Paul Scherrer Institute}, High Pressure Res. {\bf 36}, 140 (2016).\\
    \url{https://doi.org/10.1080/08957959.2016.1173690}

\bibitem{Khasanov_JAP_2022} Rustem Khasanov, {\it Perspective on muon-spin rotation/relaxation under hydrostatic pressure}, J. Appl. Phys. {\bf 132}, 190903 (2022). \\
    \url{https://doi.org/10.1063/5.0119840}

\bibitem{MUSRFIT} A. Suter and B. Wojek, {\it Musrfit: A Free Platform-Independent Framework for $\mu$SR Data Analysis}, Phys. Procedia {\bf 30}, 69 (2012).\\
\url{https://doi.org/10.1016/j.phpro.2012.04.042}

\bibitem{Amato-Morenzoni_book_2024} Alex Amato and Elvezio Morenzoni, {\it Introduction to Muon Spin Spectroscopy. Applications to Solid State and Material Sciences}, Springer Nature (2024).
\url{https://doi.org/10.1007/978-3-031-44959-8}

\bibitem{Schenck_book_1985} A. Schenck, {\it Muon Spin Rotation Spectroscopy: Principles and Applications in Solid State Physics}, Adam Hilger, Bristol, 1985.

\bibitem{Yaouanc_book_2011} A. Yaouanc and P. D. de R\'eotier, {\it Muon Spin Rotation, Relaxation, and Resonance}, Oxford Science Publications, 2011.

\bibitem{Blundell_book_2022} S. J. Blundell, R. De Renzi, T. Lancaster, F. L. Pratt (eds.) {\it Muon spectroscopy. An introduction}, Oxford: Oxford University Press, 2022.

\bibitem{Chen_arxiv_2024} K. W. Chen, X. Q. Liu, Y. Wang, Z. Y. Zhu, J. C. Jiao, C. Y. Jiang, Y. F. Guo, L. Shu, {\it Evolution of magnetism in Ruddlesden-Popper bilayer nickelate revealed by muon spin relaxation}, 	arXiv:2412.09003v1. \\
    \url{https://doi.org/10.48550/arXiv.2412.09003}

\bibitem{Pratt_JPCM_2007} F. L. Pratt, P. M. Zieli\'{}ński, M Ba{\l}anda, R Podgajny, T Wasiuty\'{n}ski and B Sieklucka, {\it A $\mu$SR study of magnetic ordering and metamagnetism in a bilayered molecular magnet}, J. Phys.: Condens. Matter {\bf 19}, 456208 (2007). \\
    \url{https://doi.org/10.1088/0953-8984/19/45/456208}

\bibitem{Sugiyama_PRB_2009} J. Sugiyama, M. M{\aa}nsson, Y. Ikedo, T. Goko, K. Mukai, D. Andreica, A. Amato, K. Ariyoshi, and T. Ohzuku, {\it $\mu +$SR investigation of local magnetic order in LiCrO$_2$}, Phys. Rev. B {\bf 79}, 184411 (2009). \\
    \url{https://doi.org/10.1103/PhysRevB.79.184411}

\bibitem{Maisuradze_PRB_2011} A. Maisuradze, Z. Guguchia, B. Graneli1, H. M. R{\o}nnow, H. Berger, and H. Keller, {\it $\mu$SR investigation of magnetism and magnetoelectric coupling in Cu$_2$OSeO$_3$}, Phys. Rev. B {\bf 84}, 064433 (2011). \\
    \url{https://doi.org/10.1103/PhysRevB.84.064433}

\bibitem{Zhang_NatCom_2020} Junjie Zhang, D. Phelan, A.S. Botana, Yu-Sheng Chen, Hong Zheng, M. Krogstad, Suyin Grass Wang, Yiming Qi, J.A. Rodriguez-Rivera, R. Osborn, S. Rosenkranz, M.R. Norman, and J.F. Mitchell, {\it Intertwined density waves in a metallic nickelate}, Nat Commun. {\bf 11}, 6003 (2020). \\
    \url{https://doi.org/10.1038/s41467-020-19836-0}

\bibitem{Khasanov_La4310_unp} Rustem Khasanov, Thomas J. Hicken, Igor Plokhikh, Vahid Sazgari, Lukas Keller, Vladimir Pomjakushin, Szymon Kr\'{o}lak, Jonas A. Krieger, Hubertus Luetkens, Tomasz Klimczuk, Dariusz J. Gawryluk, and Zurab Guguchia, {\it Pressure Effect on Density Wave Transitions in La$_4$Ni$_3$O$_{10}$}, 	arXiv:2503.04400.\\
    \url{https://doi.org/10.48550/arXiv.2503.04400}


\bibitem{Kaul_JMMM_1985} S. N. Kaul, {\it Static critical phenomenon in ferromagnets with quenched disorder}, J. Mag. Magn. Mater. {\bf 53}, 5 (1985).\\
    \url{https://doi.org/10.1016/0304-8853(85)90128-3}

\bibitem{Guillou_PRB_1985} J. C. Le Guillou and J. Zinn-Justin, {\it Critical exponents from field theory}, Phys. Rev. B {\bf 21}, 3976 (1980).\\
    \url{https://doi.org/10.1103/PhysRevB.21.3976}


    \url{https://doi.org/10.1088/0305-4470/29/17/042}

\bibitem{Liu_PRL_2023} Yu-Bo Liu, Jia-Wei Mei, Fei Ye, Wei-Qiang Chen, and Fan Yang, {\it $s\pm$-Wave Pairing and the Destructive Role of Apical Oxygen Deficiencies in La$_3$Ni$_2$O$_7$ under Pressure}, Phys. Rev. Lett. {\bf 131}, 236002 (2023).\\
    \url{https://doi.org/10.1103/PhysRevLett.131.236002}

\bibitem{Qin_PRB_2023} Qiong Qin and Yi-feng Yang, {\it High-Tc superconductivity by mobilizing local spin singlets and possible route to higher $T_{\rm c}$ in pressurized La$_3$Ni$_2$O$_7$}, Phys. Rev. B {\bf 108}, L140504 (2023).\\
    \url{https://doi.org/10.1103/PhysRevB.108.L140504}

\bibitem{Sakakibara_PRL_2024_2} Hirofumi Sakakibara, Naoya Kitamine, Masayuki Ochi, and Kazuhiko Kuroki, {\it Possible High Tc Superconductivity in La$_3$Ni$_2$O$_7$ under High Pressure through Manifestation of a Nearly Half-Filled Bilayer Hubbard Model}, Phys. Rev. Lett. {\bf 132}, 106002 (2024).\\
    \url{https://doi.org/10.1103/PhysRevLett.132.106002}

\bibitem{Shen_ChinPhysLett_2023} Yang Shen, Mingpu Qin, and Guang-Ming Zhang, {\it Effective Bi-Layer Model Hamiltonian and Density-Matrix Renormalization Group Study for the High-Tc Superconductivity in La$_3$Ni$_2$O$_7$ under High Pressure}, Chin. Phys. Lett. {\bf 40}, 127401 (2023).\\
    \url{https://doi.org/10.1088/0256-307X/40/12/127401}

\bibitem{Xue_ChiPhysLett_2024} Jie-Ran Xue and Fa Wang, {\it Magnetism and Superconductivity in the t–J Model of La$_3$Ni$_2$O$_7$ Under Multiband Gutzwiller Approximation}, Chin. Phys. Lett. {\bf 41}, 057403 (2024).\\
    \url{https://doi.org/10.1088/0256-307X/41/5/057403}

\bibitem{Yang_PRB_2023} Q.-G. Yang, D. Wang, and Q.-H. Wang, {\it Possible $s\pm$-wave superconductivity in La$_3$Ni$_2$O$_7$}, Phys. Rev. B {\bf 108}, L140505 (2023).\\
    \url{https://doi.org/10.1103/PhysRevB.108.L140505}

\bibitem{Yang_PRB_2023_2} Yi-feng Yang, Guang-Ming Zhang, and Fu-Chun Zhang, {\it Interlayer valence bonds and two-component theory for high-Tc superconductivity of La$_3$Ni$_2$O$_7$ under pressure}, Phys. Rev. B {\bf 108}, L201108 (2023).\\
    \url{https://doi.org/10.1103/PhysRevB.108.L201108}

\bibitem{Zhang_PRB_2023} Yang Zhang, Ling-Fang Lin, Adriana Moreo, Thomas A. Maier, and Elbio Dagotto, {\it Trends in electronic structures and $s\pm$-wave pairing for the rare-earth series in bilayer nickelate superconductor $R_3$Ni$_2$O$_7$}, Phys. Rev. B {\bf 108}, 165141 (2023).\\
    \url{https://doi.org/10.1103/PhysRevB.108.165141}

\bibitem{Luo_npjQuant_2024} Zhihui Luo, Biao Lv, Meng Wang, W\'{e}i W\'{u}, Dao-Xin Yao, {\it  High-Tc superconductivity in La$_3$Ni$_2$O$_7$ based on the bilayer two-orbital t-J model}, npj Quantum Mater. {\bf 9}, 61 (2024).\\
    \url{https://doi.org/10.1038/s41535-024-00668-w}

\bibitem{Fan_PRB_2024} Zhen Fan, Jian-Feng Zhang, Bo Zhan, Dingshun Lv, Xing-Yu Jiang, Bruce Normand, and Tao Xiang, {\it Superconductivity in nickelate and cuprate superconductors with strong bilayer coupling}, Phys. Rev. B {\bf 110}, 024514 (2024).\\
    \url{https://doi.org/10.1103/PhysRevB.110.024514}

\bibitem{Heier_PRB_2024} Griffin Heier, Kyungwha Park, and Sergey Y. Savrasov, {\it Competing $d_{xy}$ and $s\pm$ pairing symmetries in superconducting La$_3$Ni$_2$O$_7$: LDA + FLEX calculations}, Phys. Rev. B {\bf 109}, 104508 (2024).\\
    \url{https://doi.org/10.1103/PhysRevB.109.104508}

\bibitem{Jiang_ChinPhysLett_2024} Kun Jiang, Ziqiang Wang, and Fu-Chun Zhang, {\it High-Temperature Superconductivity in La$_3$Ni$_2$O$_7$}, Chin. Phys. Lett. {\bf 41}, 017402 (2024).\\
    \url{https://doi.org/10.1088/0256-307X/41/1/017402}

\bibitem{Lechermann_PRB_2023} F. Lechermann, J. Gondolf, S. B\"{o}tzel, and I. M. Eremin, {\it Electronic correlations and superconducting instability in La$_3$Ni$_2$O$_7$ under high pressure}, Phys. Rev. B {\bf 108}, L201121 (2023).\\
    \url{ https://doi.org/10.1103/PhysRevB.108.L201121}

\bibitem{Boetzel_PRB_2024} Steffen B\"{o}tzel, Frank Lechermann, Jannik Gondolf, and Ilya M. Eremin, {\it Theory of magnetic excitations in the multilayer nickelate superconductor La$_3$Ni$_2$O$_7$}, Phys. Rev. B {\bf 109}, L180502 (2024). \\
    \url{https://doi.org/10.1103/PhysRevB.109.L180502}

\end{thebibliography}
\end{document}